\documentclass[preprint2]{aastex}

\def\gsim{\mathrel{\raise.5ex\hbox{$>$}\mkern-14mu
             \lower0.6ex\hbox{$\sim$}}}
\def\lsim{\mathrel{\raise.3ex\hbox{$<$}\mkern-14mu
             \lower0.8ex\hbox{$\sim$}}}
\def\etal{et al.}
\def\mcg{MCG -06-30-15}

\slugcomment{Accepted for publication in the Astrophysical Journal}

\shorttitle{Fourier spectroscopy of \mcg}
\shortauthors{Papadakis, Kazanas \& Akylas}
 
\begin{document}

\title{Fourier Resolved Spectroscopy of the XMM-Newton Observations
of \mcg}

\author{I. E. Papadakis\altaffilmark{1,2}, Demosthenes 
Kazanas\altaffilmark{3}, A. Akylas\altaffilmark{4}} 

\altaffiltext{1}{Physics Department, University of Crete, Heraklion,
71003, Crete, Greece; jhep@physics.uoc.gr}
\altaffiltext{2}{IESL, FORTH, 710 03, Heraklion, Crete, Greece}
\altaffiltext{3}{Laboratory for High Energy Astrophysics, 
NASA Goddard Space Flight Center, 
Code 661, Greenbelt, MD 20771, USA; kazanas@gsfc.nasa.gov}
\altaffiltext{4}{Institute of Astronomy \& Astrophysics, National 
Observatory of Athens, I. Metaxa B. Pavlou, Penteli, Athens, 15236, Greece; 
th@astro.noa.gr}
\email{aastex-help@aas.org}

\begin{abstract}

We study the {\sl Frequency Resolved Spectra} of the Seyfert galaxy \mcg\  
obtained during two recent XMM-Newton observations. Splitting the Fourier
spectra in soft ($E \lsim 2$ keV) and hard ($E \gsim 2$ keV) bands, we find
that the soft band has a variability amplitude larger than the hard one on time 
scales longer than 10 ksec, while the opposite is true on time scales shorter
than 3 ksec. Both the soft and hard band spectra are well fitted  by power laws
of different indices. The spectra of the hard band become  clearly softer as
the Fourier Frequency decreases from $7 \times  10^{-4}$ Hz to $10^{-5}$ Hz,
while the spectral slope of the  soft band power law component is independent
of the Fourier frequency. The well known broad Fe K$\alpha$ feature is  absent
at all frequency bins;  this result implies that this feature is not  variable
on time scales shorter than $\sim 10^5$ sec, in agreement with recent line
variability studies. Strong spectral features are also  present in the
soft X-ray band (at $E \simeq 0.7$), clearly discernible in all Fourier
Frequency bins. This fact is consistent with the assumption that  they are due
to absorption by intervening matter within the source.

\end{abstract}

\keywords{ galaxies: active ---  galaxies: Seyfert  --- X-rays: galaxies}

\section{Introduction}

The ``standard" picture of the Active Galactic Nuclei (AGN) accretion  disk
arrangement consists of a geometrically thin, optically thick disk  (that
produces the quasithermal optical--UV feature known as the ``Blue  Bump")
supplemented by an overlying, hot ($\sim 10^8$ K) corona  responsible for the
observed X-ray emission. Both these components are  presumed to extend to the
innermost stable particle orbits associated  with the accreting black hole. 
This latter fact led to the suggestion that studies of the spectral and timing
characteristics of these components could provide a probe of the regime of
strong gravitational physics.

It has been argued that the reprocessing of X-ray radiation on the much cooler
accretion disk would produce a relativistically broadened,  asymmetric, Fe
K$\alpha$ fluorescence feature at 6.4 keV (Fabian \etal\  1989; Stella 1990)
and a spectral hardening of the spectrum at energies $E \gsim 10$ keV  due to
reflection of the coronal X-rays by neutral matter on the accretion  disk
surface. It was further argued that the precise shape, EW and variability
properties of this feature could allow the mapping of the space--time geometry
in the black hole vicinity (Reynolds \etal\ 1999) and/or the geometric
arrangement of the disk and the X-ray emitting plasma (Nayakshin \& Kazanas
2001).

AGN observations appear to corroborate these considerations: A broad  feature
at the correct energy was indeed identified in the ASCA spectra  of many AGN
(Nandra \etal\ 1997b). One of the broadest such asymmetric  emission features
was detected in the energy spectrum of the Seyfert 1 galaxy \mcg\ (Tanaka
\etal\ 1995). The presence of this emission feature  has been confirmed by
repeated observations of the source by all recent X-ray missions, including
Chandra (Lee \etal\ 2002) and XMM-Newton (Wilms  \etal\ 2001; Fabian \etal\
2002) and thought to suggest, on the basis of  its width, the presence of an
extreme Kerr geometry for the accreting  black hole. At energies below 2 keV
\mcg\ shows great spectral complexity attributed to absorption by partially 
ionized material and possibly also dust (Turner \etal\ 2003 and references 
therein); alternatively, these same features have also been interpreted  by
some authors due to relativistically broadened soft X-ray  emission  features
(Branduardi-Raymont \etal\ 2001; Sako \etal\ 2003).

It should be born in mind though that, however convincing, spectral  features
by themselves generally provide information only about the  properties of the
emitting (or absorbing) plasma along the observer's  line of sight, typically
just its column density. However, the size or the density of the  plasma,
parameters needed to  determine its dynamics, require additional independent
information that,  in the absence of sufficient spatial resolution, is provided
by the  sources' variability (Kazanas, Hua \& Titarchuk 1997). Hence, a more 
comprehensive approach should involve the use of both spectral and timing 
information in a combined analysis. A novel approach in this direction  has
been taken by Revnivtsev, Gilfanov \& Churazov (1999) who used the  combined
RXTE variability - spectral information to produce the energy  spectra of
Cygnus X-1 in its hard state for different Fourier  frequencies. This study
showed that the Cyg X-1 spectra depend strongly on the  Fourier frequency, with
the Fe K$\alpha$ line and the reflection component  becoming increasingly
visible with decreasing frequency. Very similar  results were obtained by the
same authors for the Galactic source GX 339-4 (Revnivtsev, Gilfanov \& Churazov
2001), while Cyg X-1 in its soft  state showed much smaller dependence of its
spectrum and Fe K$\alpha$  line on Fourier frequency (Revnivtsev, Gilfanov \&
Churazov 2000).

In the present paper we apply the method used by Revnivtsev \etal\ (1999)  to
\mcg. We chose this source because, apart from being considered { as the AGN
with the archetypal Fe K$\alpha$ profile}, it shows large amplitude variations 
on both short and long time scales (e.g.~ McHardy et al. 2005), it is bright 
in X-rays and has been the target of numerous  X-ray observations. Of
particular  interest amongst them are those by XMM-Newton  due to the large
area of its instruments, which offer high quality, low  noise spectra, albeit
at energies less than 10 keV. In \S 2 and \S 3 we discuss  the data sets and the
methodology used and we present our results in the  form of energy spectra at
three different Fourier frequencies. In \S 4  our results are discussed briefly
in the context of the standard and  other models presently in the literature.

\section{Data Reduction and Analysis Method}

\begin{figure}
\epsscale{.80}
\plotone{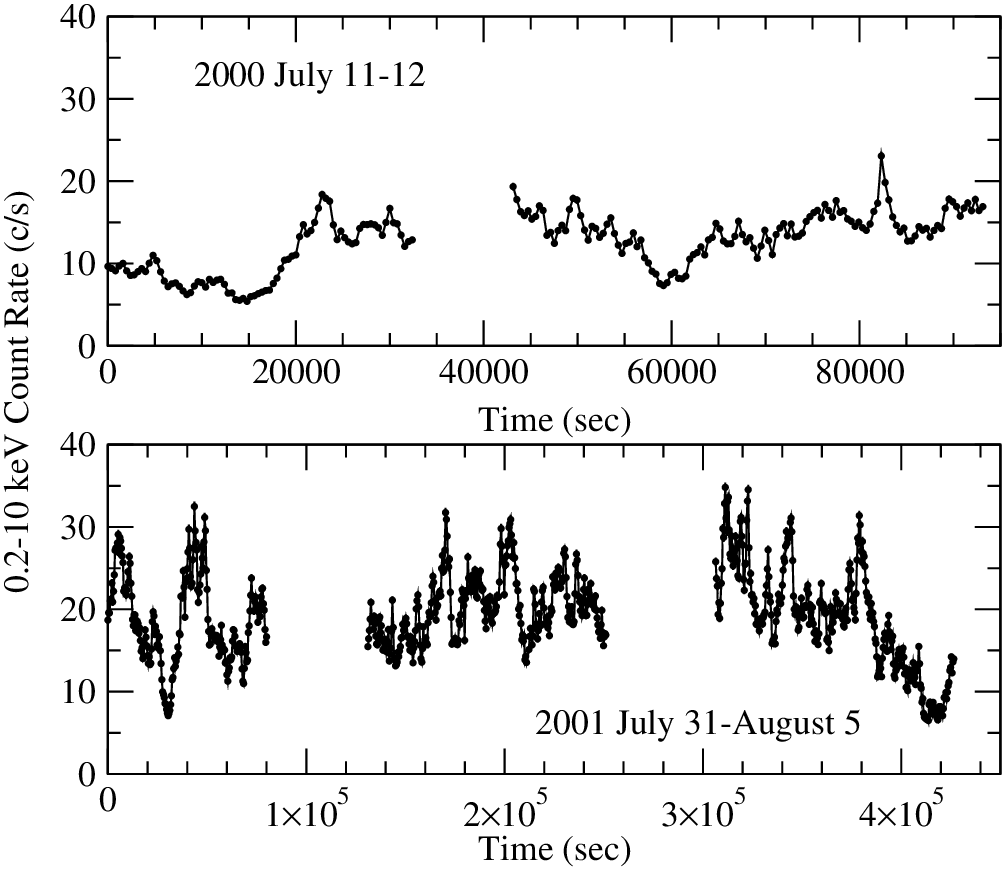}
\caption{Full-band (0.2-10 keV) EPIC-PN light curves in 400 s bins of both 
the 2000 and 2001 observations. Errors are plotted, but are very small 
(typically of the order of $\sim 0.35$ and $0.43$ counts/s,respectively).}
\end{figure}

\mcg\ has been observed with XMM-Newton twice. The first observation was 
performed on 2000 July 10-11 (revolution 108) and the second on 2001 July  31 -
August 5 (revolutions 301-303). In this work we use data from the  EPIC-PN
detector only. In both cases the source was observed on-axis and  the PN camera
was operated in small window mode, with medium filter  applied. We processed
the data using SAS v6.0.0. Source data (single and  double pixel events, i.e.
patterns $0-4$)  were extracted from circular  regions of radius
$40^{\prime\prime}$, and background events from a  source free area 2 times
larger than the source extraction area. The  background was in general low and
stable throughout the observations,  with the exception of the final few ks of
each revolution where the  background count rate increases. We kept 83 and 320
ks of ``good"  exposure time data from the 2000 and 2001 observations,
respectively.

Fig.~1 shows the 0.2-10 keV, background subtracted, 400-s binned, EPIC-PN 
light curves. The source shows large amplitude variations at all time  scales.
The raw energy spectrum, using data from both observations, is shown in the
upper panel of Fig.~2. The energy resolution is  significantly reduced for
reasons that we explain in the following  section. The photon count spectrum is
estimated in 22 energy bands only:  we consider 6 bands from 0.2 to 0.8 keV
with $\Delta E=0.1$ keV, the  bands $0.8-1$, $1-1.3$, $1.3-1.6$ and $1.6-2$
keV, 9 bands from 2 to 7  keV with $\Delta E=0.5$ keV, and finally the bands
$7-8$ and $8-10$ keV.

The main features are clearly evident even in this low resolution  spectrum.
The solid line in the upper panel of Fig.~2 shows a power law  model spectrum
of $\Gamma=2$ and Galactic absorption only  ($N_{H}=4.06\times
10^{20}$cm$^{-2}$; Elvis, Wilkes, \& Lockman 1989).  The model normalization is
adjusted to produce total number of counts  equal to those of the observed
spectra (this is the case for the normalization of all model spectra presented
in the rest of the paper). In the lower panel of Fig.~2 we  show a plot of the
data over the model ratio (filled circles). The  significantly asymmetric,
broad iron line in the energy band $\sim 5-7$  keV and strong residuals in the
soft X-ray band can be clearly seen.

\subsection{Fourier Resolved Spectroscopy}

We now discuss briefly the technique of ``Fourier resolved spectroscopy" 
introduced by Revnivtsev \etal\ (1999). Suppose we have light curves of a 
given source at different energy bands $E_{j}$, $j=1,2,\ldots M$. Let us 
denote with $x(t_{i},E_{j})$ the observed count rate at time $t_{i}$ in  the
energy band $E_{j}$ ($i=\Delta t, 2\Delta t, ..., N\Delta t$, $N$ is  the total
number of points, and $N\Delta t=T$ is the length of the light  curve). The
power spectrum (PSD) of this light curve can be estimated as  follows:

\begin{equation}
P(\nu_{k},E_{j})=\frac{2\Delta t}{N}|a_{k}|^{2},
\end{equation}

\noindent in units of (counts s$^{-1})^{2}$ Hz$^{-1}$, where 
$\nu_{k}=k/N\Delta t$ $(k=1,2,...,(N-1)/2)$ and $a_{k}=\sum_{i} 
x(t_{i},E_{j})e^{i2\pi\nu_{k}t_{i}}$. The finite Fourier  representation of the
observed time series $x(t_{i},E_{j})$ is equal to  the sum of sinusoidal terms
(``variability components") with frequencies  $\nu_k$. Their amplitudes,
$A_{k}$, are related to $|a_{k}|$  through the relation $A_{k}=(2/N)|a_{k}|$.
Using equation (1), this  relation can be rewritten as follows:

\begin{equation} 
A(\nu_{k},E_{j})=\sqrt{\frac{2P(\nu_{k},E_{j})}{N\Delta
t}}~{\rm counts~ s}^{-1}. 
\end{equation} 
$A(\nu_{k},E_{j})$, viewed as a function of $E_j$, represents the  energy
spectrum of the component with frequency $\nu_k$.

Although \mcg\ is a bright Seyfert 1 galaxy it is not possible to study  its
energy dependent variability behavior using the full energy  resolution offered
by XMM-Newton. Instead, we have split the 0.2-10 keV  energy range in the 22
bands referred to in the previous section. These  bands were so chosen to: a)
be sufficiently broad to yield light curves  of reasonably high signal-to-noise
for an accurate determination of the  power spectrum, and b)  their number in
the traditional ``soft" ($<2$  keV) and ``hard" ($>2$ keV)  X-ray channels be
roughly equal. The two  XMM-Newton observations resulted in 5 segments during
which the source  was observed almost continuously for $T\sim 32 - 120$ ks (see
Fig.~1).  For each segment we constructed background-subtracted light curves in
the  22 energy bands using a bin size of $\Delta t=100$ s. In this way a total 
of 105 light curves were produced. We used equation (1) to estimate their 
power spectrum, after subtracting the associated (constant) Poisson power. In
principle, using equation (2), we can now compute the amplitudes
$A_{l}(\nu_k,E_j),  l=1,\dots,5$. The mean of the five  $A_{l}(\nu_k,E_j)$
estimates at each $\nu_k$, viewed as a a function of $E_j$, can constitute the
``low-resolution" energy spectrum of the  variability components with frequency
$\nu_k$.

However, the energy spectra of the individual $\nu_k$'s turned out to be 
very  noisy. In fact, at high frequencies, after the subtraction of the Poisson
noise component, many of the $P(\nu_{k},E_{j})$'s are negative, thus preventing
the estimation of the respective $A(\nu_{k},E_{j})$. For this reason,  we
decided to consider three broad frequency bands : a) $8.3\times 10^{-6} -
10^{-4}$ Hz, b) $1 - 3\times 10^{-4}$ Hz,  and c) $3 - 7 \times 10^{-4}$ Hz
(hereafter ``low/LF", ``medium/MF"  and  ``high-frequency/HF" range,
respectively). These frequency bands correspond to time scales of $0.12-1.4$
days, $3-10$ ksec, and $1.5-3$ ksec, respectively. There are $N_{\rm PSD}=39$,
81  and 161 $P(\nu_{k},E_{j})$ estimates in each frequency bin. We computed
their mean, $\overline{P(E_{j})}=\sum_{l,k} P(\nu_{k},E_{j})/N_{\rm PSD}$, and
used equation (2) to estimate the average amplitude, $\overline{A(E_{j})}$, of
the variability components in each frequency band. These values comprise our
final estimate of the energy spectrum of the low, medium and  high-frequency
variability components.

\begin{figure}
\plotone{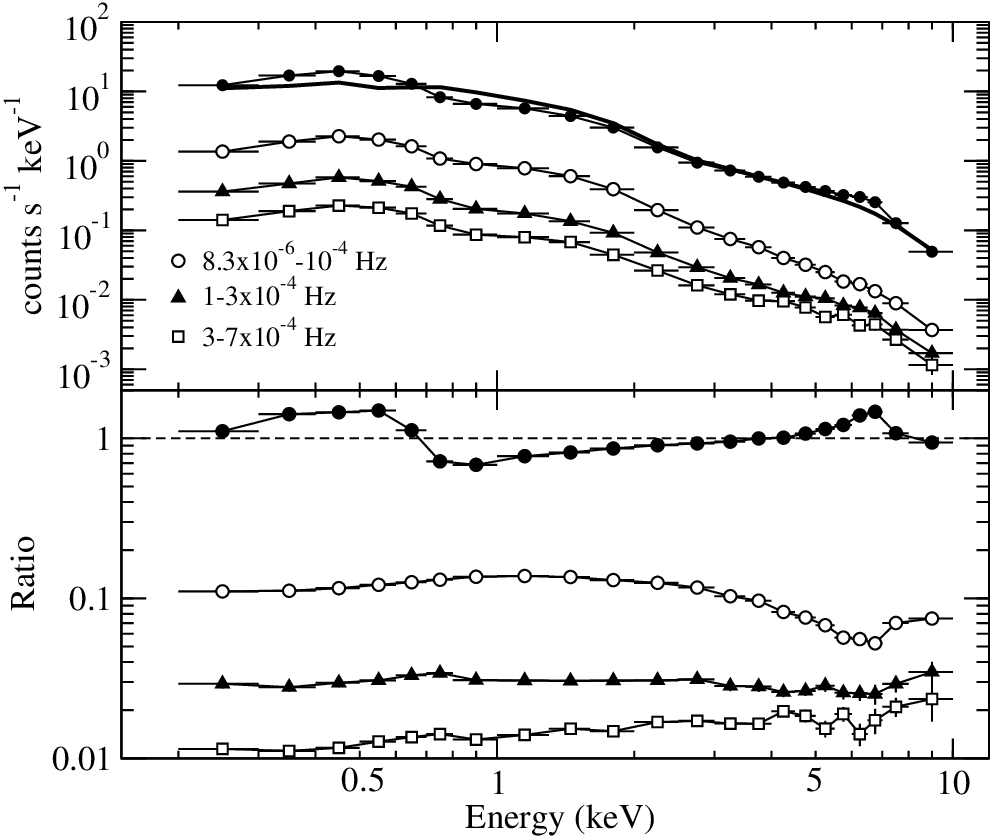}
\caption{Upper panel: Total EPIC-PN count spectrum (filled circles) and
the LF, MF and HF Fourier energy spectra (open circles, filled triangles,
and open squares, respectively). The solid line shows a $\Gamma=2$ power
law spectrum affected by the Galactic absorption only. Lower panel: Ratio
of the total spectrum to the simple power law model (filled circles) and
the Fourier resolved spectra normalized to the total EPIC-PN spectrum.}
\end{figure}

\section{Results}

The raw, count energy spectra of the LF, MF and HF components (i.e.~
$\overline{A(E_{j})}$ divided by $\Delta E$) are also shown in the  upper panel
of Fig.~2 ({ given respectively by the open circles, filled triangles, and
open  squares}). In order to gain some insight into their broad-band shape, we
divided them by the raw total energy spectrum of the source. These ratios
(shown in the lower panel of Fig.~2) should factor  out the broad instrumental
response that distorts the spectra and, as a  result, should give a better view
of their true shape, albeit only in  relation to that of our entire data set.

At energies $E < 3$ keV the LF and MF normalized spectra are consistent  with
constant, yielding respectively $\chi^{2}=14.4$ and 8.7/11 dof when  fit as
such. This indicates that the spectra of the LF and MF components  are not
significantly different in shape from the total energy spectrum  of the source.
At higher energies, the LF and HF normalized spectra  suggest that the
respective frequency spectra are significantly softer  and harder,
respectively, than the total energy spectrum. In fact, the HF  normalized
spectrum suggests that the spectrum of the high frequency  components is
systematically harder in the entire $0.2-10$ keV band.  Also, a clear decrease
of the $5-7$ keV band flux is observed in the LF  normalized spectrum. Such a
feature should be expected if there was no  significant iron line emission in
the LF spectrum. A similar feature may  also be present in the other two
normalized spectra, although not as  clear.

Irrespective of their intrinsic shape, the Fourier count  spectra can be
used to compute the amplitude of the various variability components in any
given energy band. The integral of the Fourier spectrum  over a particular energy
band (say from $E_1$ to $E_2$) yields the contribution of the variability
component with frequency  $\nu_k$ to the variance of the light curve in this
energy band. Consequently, the ratio of this integral over the integral  of the
total energy spectrum (in the same band) corresponds to the fractional root
mean square amplitude ($f_{rms})$ of this component in this energy band.

In our case, we computed the sums of $\overline{A(E_{j})}$ over the soft and
hard energy bands for the three Fourier spectra, and divided them by the
integral of the total spectrum in the same energy bands. In the soft band, we
find $f_{rms}=12.4\pm 0.5$\%, $3.05\pm 0.06$\%, and $1.30\pm 0.05$\%  for the
LF, MF and HF components, respectively. The respective values in the hard band
are: $9.7\pm 0.6$\%, $2.90\pm 0.06$\%, and $1.74\pm 0.05$\%. These results show
that on long time scales ($0.12-1.4$ days) the amplitude of the observed
variations in the soft band is larger than that in the hard band.  The opposite
holds true on short time scales ($1.5-3$ ksec). On intermediate time scales
($3-10$ ksec), the soft and hard band variability amplitudes are comparable.

\subsection{Model fits to the high energy band}  

\begin{figure}
\plotone{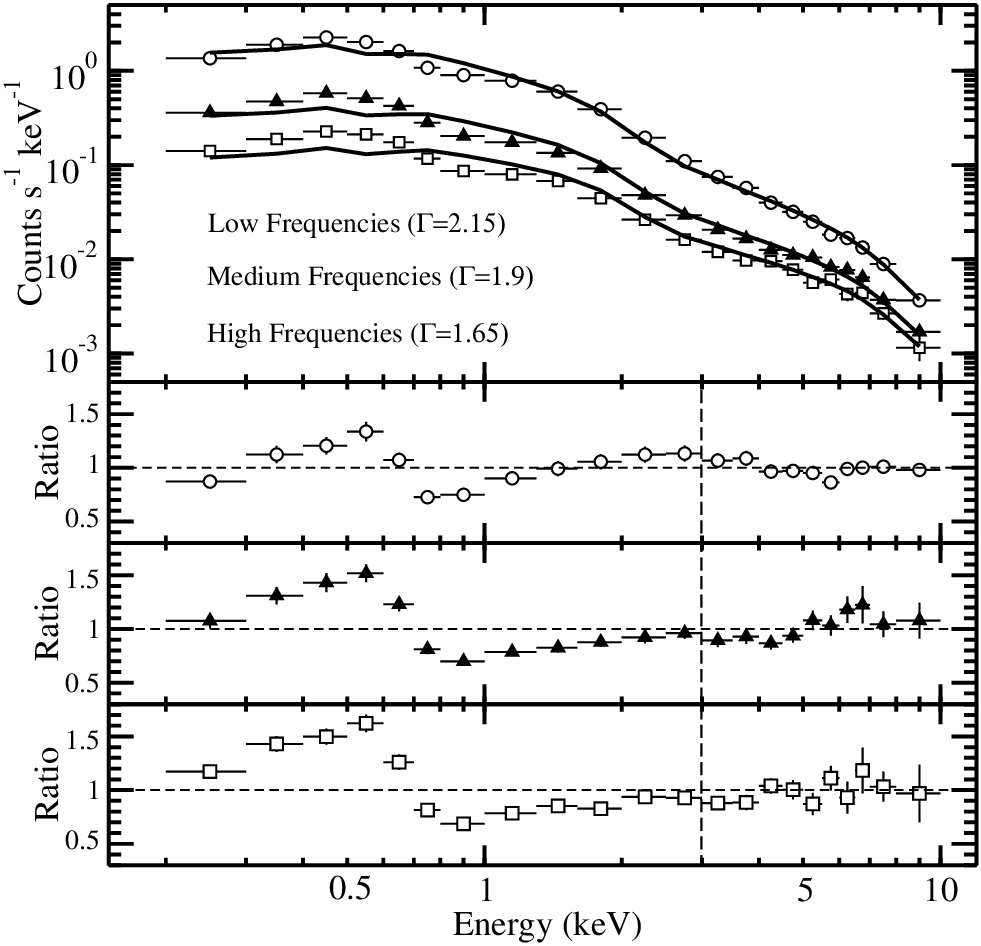}
\caption{The LF, MF and HF Fourier energy spectra (open circles, filled
triangles and open squares, respectively) plotted together with the
``best-fitting" power-law models in the $>3$ keV band (solid
lines; see text for details). The lower panels show the ratio of the
Fourier spectra to the best fitting model.}
\end{figure}

In order to get additional insight into the broad band energy spectra of  the
different frequency components we produced count spectra for power law models
with only Galactic absorption and slope increasing from  $\Gamma=1.5$ to
$\Gamma=2.5$ with a step of $\Delta\Gamma=0.05$.  The  power law normalizations
were adjusted so that the total counts in the  model and the LF, MF and HF
energy spectra be equal (the appropriate  auxiliary files for the creation of
the model spectra were produced using  the RMFGEN and ARFGEN tasks of SAS). The
upper panel in Fig.~3 shows the  LF, MF and HF energy spectra along with the
model spectrum that fits each  ``best" at $E > 3$ keV, i.e.~gives the minimum
sum of squared residuals  (weighted by the data errors square) among the model
spectra considered.  The lower three panels in the same Figure show the ratio
of the data to  the best fitting model. Our model fitting results show that the
spectral  slope decreases from the LF to the HF band. We find that above 3 keV 
the LF, MF and HF spectra are well described by a power law model with 
$\Gamma=2.15, 1.9$, and 1.65, respectively. When we fit the data/model  ratios
above 3 keV with a constant of value 1 we find $\chi^{2}=11.6, 15.1,$ and
11.2/10 dof, respectively.  We conclude that there is no strong evidence for
the presence of an iron line in any of the three  Fourier spectra.  
Furthermore, although the hard power law component is variable on all time
scales (i.e. all three Fourier frequency bands),  its slope does not remain
constant. For example, there is a significant difference of $\Delta \Gamma=0.5$
in the hard band power law slopes of the LF and HF components. In other words,
we find that the high frequency component has harder spectrum than that of the
lower frequency one.

\subsection{Model fits to the full band energy spectrum}

At energies below $\sim 2$ keV, a simple power law model with Galactic 
absorption does not fit well the spectra of any of the three Fourier frequency
bands.  In all cases, the plot of the data/model ratio as a function of energy
is qualitatively similar to the respective plot in the case of the power law
model fit to the total energy spectrum (shown on the top of the lower panel in
Fig.~2): we observe a strong,  broad absorption feature at energies $\sim
0.7-2$ keV, and a broad ``hump"  at lower energies, whose amplitude appears to
increase with increasing frequency. We modeled the soft X-ray band spectrum 
both for the entire data set and the three Fourier components as follows.

Since we have undersampled the energy resolution of the EPIC-PN data, it is
possible to model the soft X-ray complex features considering simple,
phenomenological  components such as power laws and absorption edges (as
opposed  to constructing physical models of the absorber and the emitting
source). To this end, following Turner \etal\ (2003), we first considered the
entire energy spectrum being the sum of two power laws, one for each of the
soft and hard bands, joining at an energy $E_{\rm break}=1$ keV and with
$\Gamma_{\rm soft}>\Gamma_{\rm hard}$. Furthermore, in addition to  Galactic
neutral absorption, we also considered the possibility of intrinsic cold
absorption as well. We produced count spectra keeping $\Gamma_{\rm hard}$ fixed
to the value 2, and  $\Gamma_{\rm soft}$ increasing from 2.05 to 3 in steps  of
$\Delta\Gamma=0.05$. In each step, we produced three different spectra with
$N_{\rm H, intr}=10^{20}, 10^{21},$ and $10^{22}$ cm$^{-2}$. We found that the
model spectra with $\Gamma_{\rm soft}\sim 2.4-2.6$ and $N_{\rm H,
intr}=10^{20}$ cm$^{-2}$ agree ``best" with the observed spectrum. 

We then produced a new set of model spectra with $\Gamma_{\rm soft}=2.4-2.6$,
and  $N_{\rm H, intr}=10^{20}-10^{21}$ cm$^{-2}$ (using $\Delta\Gamma=0.05$
and  $\Delta N_{\rm H, intr}=10^{20}$ cm$^{-2}$, respectively). Following Wilms
\etal\ (2001), at each spectrum corresponding to a given pair of $(\Gamma_{\rm
soft}, N_{\rm H, intr})$ values we also added absorption edges at 0.74, 1 and
1.36 keV (these threshold energies correspond with the expected absorption
edges of  O{\small VII}, Ne{\small IX}/Mg{\small X}, and Ne{\small X}). As for
the optical depth, we used values of $\tau =0.5-1$ for the first edge, and
$\tau =0-0.5$ in the  other two cases, with a step of $\Delta\tau=0.1$. We then
compared each $(\Gamma_{\rm soft}, N_{\rm H, intr}, \tau_{0.74 keV},  \tau_{1
\rm keV}, \tau_{1.36 \rm keV})$ model with the observed spectrum in order to
find the one with the minimum $\chi^2$.

\begin{figure}
\plotone{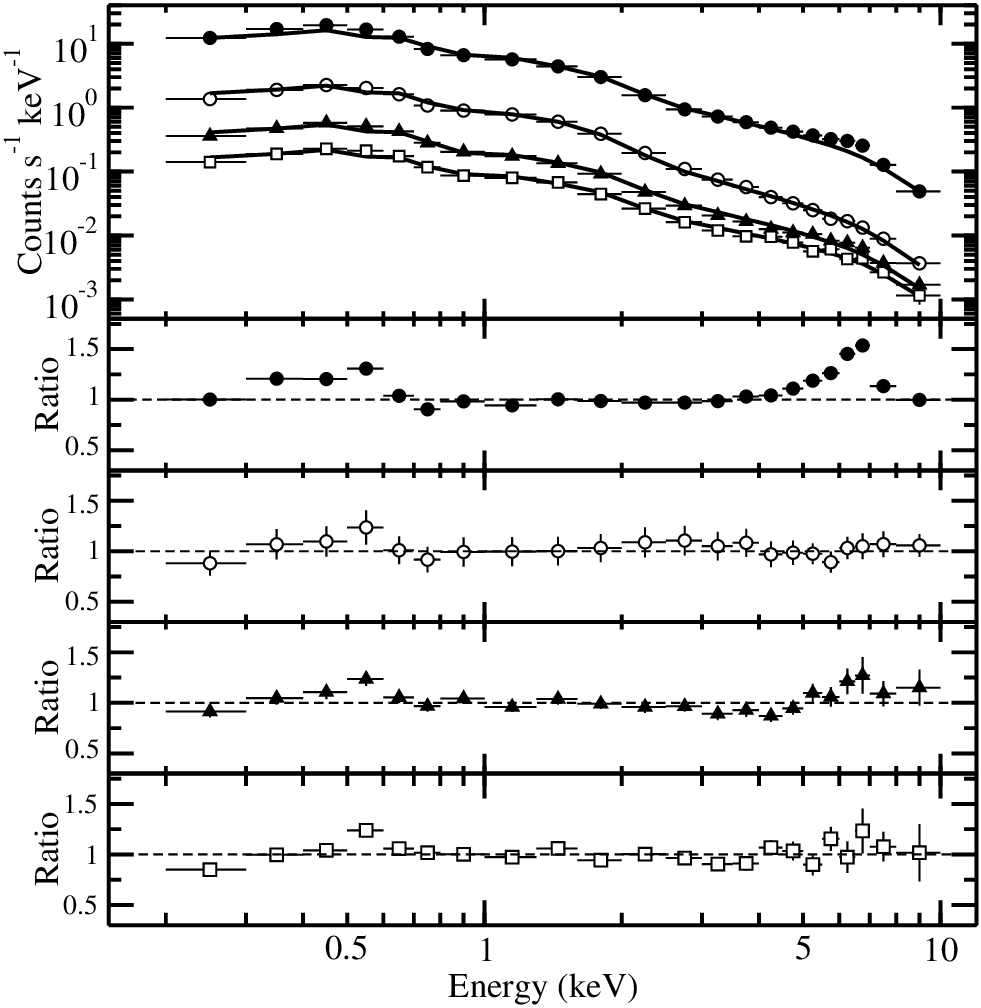}
\caption{The total, LF, MF and HF Fourier energy spectra (filled circles, 
open circles, filled triangles and open squares, respectively) plotted 
together with the ``best-fitting" model discussed in Section 2.2.2. The 
lower panels show the ratio of the spectra to the respective best fitting 
model.}
\end{figure}

The filled circles on the upper panel of Fig.~4 show the total energy spectrum
of the source, together with our ``best fitting model": $\Gamma_{\rm soft}=2.5,
N_{\rm H, intr}=3\times 10^{20}$ cm$^{-2}$, and  $\tau_{0.74 \rm keV}=0.7,
\tau_{1 \rm keV}=0,$ and $\tau_{1.36 \rm keV}=0.2$. The data/model ratio is
plotted on the second panel of the same Figure. The model describes rather well
the overall shape of the spectrum, except in the vicinity of  the Fe K$\alpha$
line and the $0.3-0.6$ energy band, where a strong `hump' is evident. As
discussed earlier, this feature has been attributed to relativistically
broadened K$\alpha$ emission from oxygen  (Branduardi-Raymont  \etal\ 2001,
Sako \etal\ 2003). Another possibility is that the gas responsible for the
extra absorption within \mcg\ may not be cold but partially ionized. In this
case, the depth of the neutral oxygen edge at $\sim 0.54$ will be smaller,
reducing the discrepancy between model and observed spectra in this energy band
(Turner \etal\ 2003). 

Based on the results of the best fit procedure used to model the total
spectrum,  we examined whether a similar procedure could provide acceptable
fits to  the spectra of the individual Fourier components as well. The open
squares in the upper panel of Fig.~4 correspond to the HF Fourier spectrum
while the solid line shows the best fitting  ``broken power law with
intrinsic absorption and two edges" model. The model parameter values are
listed in Table 1.  The bottom panel in the same Figure shows the data/model
ratio. Apart from the point in the $0.5-0.55$ keV band, the model fits the
spectrum quite well ($\chi^{2}=24.6/21$ dof). A similar model  (without the
presence of an edge at 1.36 keV) fits reasonably well the LF spectrum as well
($\chi^{2}=17.4/21$ dof when we omit the point at $\sim 0.5$ keV; the
normalized residuals in this case are plotted in the third panel from top in
Fig.~4, and the best fitting model parameters are listed in  Table 1). In
the case of the MF spectrum, the addition of an extra edge at 0.85 keV (the
threshold energy of the O {\small VIII} edge) is necessary to provide a good
fit ($\chi_{2}=26.7/21$ dof, if the $\sim 0.5$ point is again not taken into
account; see second panel from bottom in Fig.~4 for the data/model ratio plot
in this case, and Table 1 for the model fitting results).

In summary, the overall spectrum of the source is well described by two power
law components (one for the low energy and one for the high energy sections of
the spectrum), intrinsic absorption, as well as two absorption edges. These
distinct power law components are present in the spectra of the individual
Fourier components as well, indicating that they are variable. However,
contrary to the hard band power law, $\Gamma_{\rm soft}\sim 2.5$, independent of
the Fourier frequency, suggesting a different physical origin between these two
components. Obviously, this result is somehow model dependent. Since the broken
power law is a simple phenomenological model, there is no reason to exclude an
alternative parametrization of the observed spectra, which may lead to
different results, regarding the variability properties of the soft excess.

Finally, absorption features are evident in the spectra of all three Fourier
components. The $\sim 0.7$ keV absorption edge appears to be equally strong
(with $\tau\sim0.7$) in all Fourier spectra. An edge at $\sim 1.36$ keV 
($\tau\sim 0.2-0.3$) is evident in the MF and HF spectra, and a third edge (at
$\sim 0.85$ keV; $\tau\sim 0.25$) is present in the MF spectrum only. The
differences in the presence of edges in the three  Fourier spectra could be due
to the  fact that the MF spectrum has the highest signal to noise ratio among
them. For example, the addition of two edges at 0.85 and 1.36 keV (with
$\tau=0.25$ and 0.2, respectively) in the LF best fitting model is acceptable
at a confidence level of  $\sim 10$\%, while the addition of a 0.85 keV edge
($\tau=0.25$) in the HF best model fitting spectrum provides an acceptable fit
at the $\sim 5$\% level. 

\section{Discussion, Conclusions}

We have applied the Fourier frequency resolved spectral analysis method  to the
XMM-Newton data of \mcg. This work represents the first application of this
method to the X-ray data of an AGN. Our main results are the following: a) The
soft and hard bands of the spectrum of \mcg\ exhibit different RMS variability
at different Fourier frequencies (time scales). b) Both the soft and 
hard band power law components are variable on  all time scales. However,
while  the hard band power law becomes  progressively steeper with decreasing
Fourier frequency by  $\Delta\Gamma\sim 0.25$ between each of the HF, MF and LF
bands, no spectral slope variations are observed in the soft band power law
component. c) An  iron line, while present and clearly broad and asymmetric in
the energy spectrum of the entire observation of this source, it is apparently
absent in the energy spectra of the HF, MF and LF bands. d) A significant
edge-like feature at $E  \simeq 0.7-2$ keV is present in the spectra of all
frequency bands.

Past studies based on ``excess variance" analysis of $\sim 1$ day long light
curves have shown that the variability amplitude in AGN is higher in the soft
than the hard band (Nandra \etal\ 1997a, Leighly 1999). However the variance, 
being the integral of the power spectrum over frequency, say from $\nu_{\rm min}$ to
$\nu_{\rm max}$, does not provide information on the variability  amplitudes at
the different time scales. This information, on the other  hand, is provided by
the Fourier resolved spectra obtained in our analysis. We thus find that the
hard band variations are of larger amplitude than the soft band ones on time
scales shorter than $\sim 3$ ksec, while the opposite effect is observed on
time scales longer than 10 ksec. The distinctly different timing properties
of  these two bands, in conjunction with their very different spectra argue
that they represent emission by different physical components, likely of
different physical dimension.

However, even within the hard band itself, the spectral properties do  depend
on the Fourier frequency, as the spectra become softer with  decreasing Fourier
frequency. This can be attributed to the fact that  the hard band fast varying
components have spectra significantly  harder  than those of slower ones, or
alternatively, that the higher energy variations are consistently shorter
than those of lower energies. This result is not consistent with the suggestion
of Vaughan \& Fabian (2004) that the spectral variability of \mcg\ over the
entire $0.2-10$ keV band can be explained by the presence of a constant
component and a power law component which varies only in normalization.

On the other hand,  this  conclusion is in agreement with the power spectrum
analysis of the XMM-Newton data of \mcg\ by  Vaughan \etal\ (2003) who find
that the power spectrum becomes flatter  with increasing energy above the
``break" frequency of $\sim 10^{-4}$ Hz. Similar behavior (power spectrum
hardening at high frequencies with increasing energy) has been recently
observed in other AGN (NGC~7469: Papadakis, Nandra \& Kazanas, 2001; Mkn 766:
Vaughan \& Fabian, 2003; NGC~4051: McHardy \etal\ 2004, and 1H 0707-495, Ton
S180 with the use of structure functions, Leighly 2004), as well as in the
galactic black hole candidate Cyg X-1. This result by itself, ignoring the
constraints imposed by the presence and variability properties of the soft
band, could be interpreted as variability due to a Comptonizing corona of
decreasing  electron temperature with radius, along the general lines discussed
in  Kazanas, Hua \& Titarchuk (1997). 

The same interpretation, however, is not consistent with the variability of 
the soft component whose slope seems independent of the Fourier frequency.  At
this point, in the absence of lower energy data and based on the  general shape
of this component we are willing to speculate that it is due  to emission by a
(non-standard) accretion disk. As such we have in mind the ADIOS flows of
Blandford \& Begelman (1999). Assuming a viscosity  parameter $\alpha \simeq
0.3$, a black hole mass $M \simeq 10^6$ M$_{\odot}$ and a disk size $R \simeq
10 \; R_S$, we obtain a characteristic variability scale $t_{\rm var} \simeq
(R/c) \alpha^{-1} (R/R_S)^{3/2} \simeq 10^4$ s,  in agreement with
observations. The disk photons could then serve as the  seeds needed to produce
the harder power law component by hot electrons  in a corona interior to and/or
overlying part of this disk. In this respect, one  should bear in mind that the
lags between the soft (0.2-0.7 keV) and the hard  (2-10 keV) bands  (Vaughan
\etal, 2003) are in general agreement  with such an interpretation and the
observed tight correlation between the  light curves of the soft and hard
spectral components.

The results of our analysis indicate that the Fe K$\alpha$ line shows  no
significant variations on  time scales shorter than $\sim 1-2$ days and are in
agreement with the  results of the previous studies (e.g. Vaughan \& Fabian
2004, and  references therein). The only grounds for concern in this conclusion
is the MF energy spectrum. First of all, the residuals around $\sim 5-7$ keV 
in the ``data/best model" ratio plot (Fig.~3 and 4) are suggestive of the 
presence of a line feature (albeit of small amplitude). Furthermore, if  the MF
spectrum were consistent with a just power law of index $\Gamma =  1.9$, we
would expect a decrease in the above ratio by $\sim 40$\% in the  $5-7$ keV
band (the strength of the line above the $\Gamma = 1.9$ power  law in the total
spectrum; see top plot in the bottom panel of Fig.~2),  but we observe a
smaller drop ($\sim 15$\%), in the ``MF/total  spectrum" ratio (Fig.~2), a
value that implies a possible difference in  the true MF slope from the assumed
value of $\Gamma = 1.9$. We plan to investigate this issue further with the use
of more data from recent Chandra, RXTE and ASCA observations of the source. 

Our  conclusions are similar to those of Revnivtsev \etal\ (1999) for Cyg X-1 
in its low/hard state. They found the Equivalent Width (EW) of the iron  line
in the spectrum of the $\sim 1$ Hz and $\sim 10$ Hz Fourier  components to be
80 and 50 eV, respectively. We find the $3\sigma$ upper  limit in the EW of the
line to be $\sim 60$ eV in the spectrum of the  $\sim 10^{-5}-10^{-4}$ Hz
components in \mcg. The ratio of these  frequency bands in the two sources is
$\sim 10^5$, roughly comparable to  the ratio of the black hole mass in the two
systems, if we assume a $10$  M$_{\odot}$ and $\sim 10^{6}$M$_{\odot}$ mass for Cyg
X-1 and \mcg\ (McHardy \etal\ 2005), respectively. As noted by Revnivtsev
\etal\ (1999), the most straightforward  interpretation of the absence of a
line in the frequency resolved spectra  of frequency higher than $\nu_k$ is
that the X-ray reprocessing matter is  located at a distance $R_{\rm rep} \gsim
c \cdot [\nu_k/2\pi]^{-1}$. For  $\nu_k \simeq 10^{-5}$ Hz this is $R_{\rm rep}
\sim c \cdot 15000 ~{\rm  s} \simeq 4.5\times 10^{14}$ cm, or $1500 R_{S}$ for
a $10^{6}$  M$_{\odot}$ black hole mass. If this is indeed the case, the line
flux  light curve should be much smoother than that of the continuum, and its 
variations should be of amplitude smaller than that of the continuum, a  fact
in agreement with observations (e.g.~Vaughan \& Fabian 2004).  However, such
large distances are hard to reconcile with the observed  line width that
requires line emission size $R_{\rm rep} \lsim 10R_S$. Recently, Miniutti
\etal\ (2003) suggested that this lack of variability  could be explained by a
combination of light bending effects and vertical  motions of the X-ray source
above the accretion disk. { Alternatively, the large width of  the line maybe due
to its downscattering in an expanding wind, as suggested recently by Titarchuk,
Kazanas \& Becker (2003). Inoue \& Matsumoto (2003) also argue that the large
width of the line is an artifact caused by the fact that the continuum suffers
from absorption by various layers of warm absorbing material with variable
column density and/or covering factor on time scales $\sim 10^5$ s. In this
case, we should observe absorption features in the $5-8$ keV band of the Fourier
spectra, but none are clearly evident in the residual plots (Fig.~3 and 4).

Of particular interest is the feature at $\simeq 0.7$ keV, which is easily
discernible in the energy spectra of all 3 frequency bands. It has  been
attributed to a combination of O VII edge and FeI absorption (Turner \etal
2003). In this case one can set a limit on the distance of the absorber,
assuming that the  later remains unaffected by the changes in the continuum
flux; this  assumption is justified by the absence of a strong dependence of
this  feature's depth on the Fourier frequency. This  sets a limit to the value
of the photoionization parameter of the  absorbing medium to values $\xi = L/n
\, R^2 \lsim 30$ erg\,cm/s. On the  other hand, the depth of the feature sets
also a limit on the column  density of the absorbing gas $N_H = n \, R \simeq
10^{22}$ cm$^{-2}$,  values consistent with the detailed models of Turner
\etal\ (2003); this  results, then, in $R \simeq L / \xi \, N_H \simeq 10~{\rm
pc}~ (L_{43.5}/  \xi_{1.5} N_{H_{22}})$, where the subscripts denote the base
10 exponent  of the value of the corresponding parameter in cgs units. 

As we have already mentioned, the same feature along with the spectral
``hump"  in the residual plots of Fig.~4 around 0.5-0.55 keV has been
interpreted as a  relativistically broadened O K$\alpha$ line. The presence of
this feature in  the spectra of all Fourier components (and in particular 
those of highest Fourier frequency) is, taken at face value, consistent with
such an interpretation. Interestingly, an absorption feature by matter with
properties obtained in the previous paragraph would also conform with the same
phenomenology. On the other hand, the broad  Fe K$\alpha$ line which is modeled
as emission by plasma in the same  physical location as that responsible for an
O K$\alpha$ line, has very  different timing properties, as it is absent from
any of the Fourier resolved spectra. This difference in the timing properties
of these two components that have presumably their origin in the kinematics of
the same plasma makes the presence of a soft X-ray line  suspect. In addition,
when one  considers the effects of the highly variable hard X-ray component on
the ionization structure of the line emitting plasma that would highly ionize
the  accretion disk plasma (Nayakshin, Kazanas \& Kallman 2000; Nayakshin \&
Kazanas  2001), the interpretation of this feature as a relativistically
broadened O K$\alpha$ line becomes doubtful.

In conclusion, the Fourier resolved spectroscopy technique discussed in the
present work is a powerful instrument in analyzing the spectro-temporal
properties of AGN and accreting compact objects in general. By attributing
specific spectral properties/features to specific Fourier frequencies/time
scales it allows one to further dissect the structure of these sources. It is
hoped that the additional information provided by this technique  will allow
for a deeper understanding of this structure and the associated physics. Our
work is but a first step in this direction. At this point it is not clear
whether the properties implied by our analysis represent general AGN trends or
idiosyncrasies of \mcg. We hope that analysis of additional objects along the
same lines will help decide this question. 

\acknowledgments

DK would like to acknowledge a stimulating discussion with Tim Kallman. 
Part of this work was supported by the General Sectreteriat of Research 
and Technology of Greece.

\clearpage

\begin{table}
\begin{center}
\caption{Best fitting values of the ``broken power law plus absorption edges"
model parameters, for the LF, MF, and HF Fourier spectra (FS). The uncertainties
quoted are at the 90 per cent level for one interesting parameter. They are
multiple of the respective parameter step size used in the construction of the
various models, and correspond to the values which increase $\chi^{2}$ (when
omitting the point at 0.55 keV) by $\Delta\chi^{2}\gsim 2.71$, keeping the other
parameters fixed to their best fitting values.}
\begin{tabular}{lcccccc}
\tableline\tableline
FS & N$_{\rm H}$ & $\Gamma_{\rm hard}$ & $\Gamma_{\rm soft}$ &
$\tau_{0.74\rm keV}$ & $\tau_{0.85\rm keV}$ & $\tau_{1.36\rm keV}$ \\
   & $\times 10^{20}$ cm$^{-2}$ &  &  &  &  & \\
\tableline
LF & $3\pm1$ & $2.25\pm0.05$ & $2.4\pm0.15$ & $0.7^{+0.4}_{-0.25}$ & $-$ & $-$\\
MF & $3\pm1$ & $2^{+0.10}_{-0.05}$ & $2.45\pm0.1$ & $0.7^{+0.15}_{-0.1}$ &
$0.25\pm0.15$ & $0.3^{+0.15}_{-0.1}$ \\
HF & $3\pm1$ & $1.7\pm0.05$ & $2.5^{+0.05}_{-0.15}$ & $0.8\pm0.15$ & $-$ &
$0.2^{+0.15}_{-0.1}$ \\
\tableline
\end{tabular}
\end{center}
\end{table}

\end{document}